\documentclass{svmult}
\usepackage{theorem}
\usepackage{graphicx}
\usepackage{amssymb,amsmath}
\begin{document}
\frenchspacing
\title{Natural vs. Artificial Topologies on a Relativistic Spacetime}
\author{Kyriakos Papadopoulos }
\authorrunning{K. Papadopoulos}
\institute{K. Papadopoulos \at Department of Mathematics, Kuwait University, PO Box 5969, Safat 13060, Kuwait\\ 
\email{kyriakos.papadopoulos1981@gmail.com} }
\maketitle

\abstract{
Consider a set $M$ equipped with a structure $*$. We call a natural topology $T_*$, on $(M,*)$, the topology induced by $*$. For example, a natural topology for a metric space $(X,d)$ is a topology $T_d$ induced by the metric $d$ and for a linearly ordered set $(X,<)$ a natural topology should be the topology $T_<$ that is induced by the order $<$. This fundamental property, for a topology to be called ``natural'', has been largely ignored while studying topological properties of spacetime manifolds $(M,g)$ where $g$ is the Lorentz ``metric'', and the manifold topology $T_M$ has been used as a natural topology, ignoring the spacetime ``metric'' $g$. In this survey  we review critically candidate topologies for a relativistic spacetime manifold, we pose open questions and conjectures with the aim to establish a complete guide on the latest results in the field, and give the foundations for future discussions. We discuss the criticism against the manifold topology, a criticism that was initiated by people like Zeeman, G\"obel, Hawking-King-McCarthy and others, and we examine what should be meant by the term ``natural topology'' for a spacetime. Since the common criticism against spacetime topologies, other than the manifold topology, claims that there has not been established yet a physical theory to justify such topologies, we give examples of seemingly physical phenomena, under the manifold topology, which are actually purely effects depending on the choice of the topology; the Limit Curve Theorem, which is linked to singularity theorems in general relativity, and the Theorem of Gao-Wald type of ``time dilation'' are such examples.
}

\section{Motivation: the Topologisation Problem.}

Almost six decades from the first papers on Einstein's theory of relativity, and simultaneously with the appearance of the first results on spacetime singularities, a freshly new discussion was initiated on whether the manifold topology should be called a {\it natural} topology for a spacetime or not. A spacetime  $(M,g)$, in general relativity, is a four-dimensional, time oriented, connected, $C^d$ manifod, which is equipped with a $C^{d-1}$ Lorentz ``metric'' $g$ \footnote{The term metric, for the Lorentz tensorfield, is an abuse of language, as was also pointed by Zeeman in \cite{Zeeman1}, but it is so widely used that we will put it in quotes, in order to distinguish from the Riemann metric.} (see, for example, \cite{Penrose-difftopology}). Thus, the problem of assigning a spacetime $(M,g)$ to a natural topology should take into account the Lorentz tensor field $g$. This idea lies on the principle that if one considers a set $M$ equipped with a structure $*$, then a natural topology $T_M$ (or $T_*$), on $(M,*)$, should be induced by $*$; otherwise, such a topology cannot be called a natural topology on $(M,*)$.

A serious problem that appears when one uses the manifold topology as a natural topology $T_M$, for a spacetime $(M,g)$, is that $T_M$ is a natural topology for the manifold $M$, as it is induced by the metric structure of the manifold, but it is not natural in $(M,g)$, where $g$ is the Lorentz ``metric''. As a consequence, the manifold topology does not incorporate the causal structure of the spacetime and, under this topology, the spacetime itself carries properties that might not be as natural as we once thought to be. In section 2 we will review the obvious differences between $T_M$ and appropriate candidates for a spacetime topology and how the properties of $T_M$ are incompatible with the structure of light-cone, a structure which corresponds to each point in the spacetime. In section 3 we will mention issues related to the {\it singularity problem} in general relativity, and how the choice of an appropriate natural topology might influence the way that we view singularities. We will extend the discussion to naked singularities and the world of wormholes, all in the frame of spacetime topology. In section 4 we will see a mysterious duality between spacelike and timelike, with respect to two dual order relations in a spacetime, each of which it induces a topology which is dual, in a particular sense, to the other. This section, as well as the previous one, will give a lot of space for questions, that we will list in the concluding section 5.

 There is a general confusion of the meaning of the term ``natural'', in topology, and this has leaded to a sequence of misunderstandings in the field spacetime geometry. In a discussion like this one, a topology is not just a tool, but something vital for the description of a spacetime as a mathematical entity. It is evident that under appropriate topologies a spacetime cannot admit singularities and several other effects, including for example the Gao-Wald ``time-dilation'', which is related to a property called causal pseudo-convexity; 
such effects are a result of the exclusive use of the manifold metric topology instead of a topology which embodies the causal and conformal structure of spacetime. A finer topology, than the manifold one, might not be related in a straightforward way to the metric structure of the manifold (as it might be not metrizable), but it contains coarser topologies, such as the usual manifold metric-topology, which can do this job.

The Riemann metric itself has proven to have a significance in theories like the Wick rotation (for a critical review on this topic see for example Penrose \cite{Road-to-reality}), but a topology which is induced by the Riemann metric is far from being called natural in a spacetime. In this Chapter, we will restrict our entire discussion to general relativity, and even if we are against the use of the term ``natural'' for the manifold topology, we should highlight that an appropriate Riemannian metric will still play a significant role in the construction of (really) natural spacetime topologies, different than the manifold one.

\section{What is (or should be) the role of spacetime topology?}

In order to answer the question of the title in this section, we need first to list properties of the manifold topology $T_M$ that make it an inappropriate choice for a natural topology for a spacetime $M$. Zeeman, in  1967 (see \cite{Zeeman1}), pointed that:

\begin{enumerate}

\item The Minkowski space, $(M,g)$, has $M = \mathbb{R}^4$ and under  the Euclidean topology $T_{\mathbb{R}^4}$, on $\mathbb{R}^4$, it is locally homogeneous (in the sense that it looks, topologically, the same at any point). The Minkowski space is not just the set $M$ though; it is the pair $(M,g)$  and this is not a locally homogeneous space; at each point there corresponds a lightcone, which separates spacelike from timelike vectors.

\item The group of all homeomorphisms of $(M,g)$ under the Euclidean topology $T_{\mathbb{R}^4}$ is vast and has no known physical meaning. An appropriate topology should associate the group of homeomorphisms to the Lorentz group and dilatations.

\end{enumerate}

G\"obel, in 1976, generalised the arguments of Zeeman for curved spacetimes, highlighting that the manifold topology (the analogue of the Euclidean topology in the case of the flat Minkowski space) is artificial both in mathematical as well as in a physical sense. He added that experts  were primarily concerned with Riemannian structures, where the manifold topology is indeed natural, and not with spaces with a pseudo-Riemannian metric (Lorentz metric is a particular example). It is rather interesting the comment that G\"obel adds, that it is not plausible to consider a spacetime as locally Euclidean and there is no justification why it should be: ``There are no experiments known to justify a Euclidean topology along lightlike geodesics''.

So, Zeeman, as a solution to the problems that he pointed out, came up with a topology which mimics the Euclidean space $\mathbb{R}^4$, in the sense that it induces the $1$-dimensional Euclidean topology on $\mathbb{R}$ and the $3$-dimensional Euclidean topology on $\mathbb{R}^3$. He named this topology ``the Fine topology $F$ on Minkowski space $M$'', and defined it to be the {\em finest} topology on $M$, that induces the $1$-dimensional Euclidean topology on every time-axis and the $3$-dimensional Euclidean topology on every space-axes. Zeeman's intuition worked pretty successfully, since he proved that, under $F$, the group of homeomorphisms of $M$ is the Lorentz group with translations and dilatations, a significant result, indeed.

G\"obel (see \cite{gobel}) extended Zeeman's result to general relativity, by giving the definition of the analogue of $F$: let $M$ be a spacetime manifold, $T_M$ its manifold topology,  and let $S$ be a collection of subsets of $M$. A set $A \subset M$
is open in $Z(S,T_M)$, a topology in the class $\mathfrak{Z}-\mathfrak{G}$ of Zeeman-G\"obel, if $A \cap B$ is open
in $(B,T_M|B)$, the subspace topology of the manifold topology $(M,T_M)$
with respect to $(B,T_M)$, for all $B \in S$. The finest such topology, call it $\mathcal{F}$, is the general relativistic analogue of $F$. Under $\mathcal{F}$, and without any restrictions on the spacetime $M$, G\"obel showed that the group of all homeomorphisms of $M$ is the group of all homothetric transformations of $M$, leading to the fact that a homeomorphism, under $\mathcal{F}$, is an isometry.

Hawking, King and McCarthy (and in communication with G\"obel) in \cite{Hawking-Topology} emphasized that the standard manifold topology merely characterizes continuity properties, and proposed a topology which determines the causal, differential and conformal structures of spacetime, but criticized Zeeman-G\"obel topologies $\mathfrak{Z}-\mathfrak{G}$ of having the following disadvantages:

\begin{enumerate}

\item A three-dimensional section of simultaneity has no meaning in terms of physically plausible experiments.

\item While the isometry and conformal groups of $M$ are significant physical, it is not necessary clear that this is so for the homothecy group of $M$.

\item $F$ is technically complicated; in particular, the fact that no point has a countable neighbourhood basis makes $F$ hard to calculate with.

\end{enumerate}

We believe that point number 3, of Hawking-King-McCarthy, is not so fruitful; one cannot expect to have a natural topology (as we defined the term ``natural'' in Section 1) and simultaneously ``easy to use''; if the topology is difficult to handle with, this can be due to the complicated structure of the universe set in which the topology is defined.

The topology that Hawking-King-McCarthy proposed is widely known as the {\em Path topology} on a spacetime, and is defined as follows. For each $x \in M$ and each open neighbourhood $U$ of $x$, let $I(p,U)$ denote
the set of points connected to $p$ by a timelike path lying in $U$ and by
$K(p,U)$ the set $I(p,U) \cup \{x\}$. By choosing an arbitrary Riemannian metric
$h$ on $M$, let $B_\epsilon(x)$ denote an open ball centered at $x$ with radius
$\epsilon >0$, with respect to $h$.
The {\em path topology} $\cal{P}$, on $M$, is defined to be the finest topology such that the induced
topology on every timelike curve coincides with the topology induced from the manifold
topology. Hawking et al. proved that
the sets of the form $K(p,U) \cap B_\epsilon(x)$ form a basis for the topology $\cal{P}$, giving to $\cal{P}$ properties, like: $\cal{P}$ has an explicit neighbourhood basis, $\cal{P}$ is strictly finer than $T_M$ and incomparable to $\mathcal{F}$, the $\cal{P}$-continuous paths are Feynman paths (for proofs of these statements see \cite{Hawking-Topology}) and overall advantages like: $\cal{P}$ determines both the causal, differential as well as conformal structure of $M$, making calculations linked to these structures purely topological.

Low has shown that the Limit Curve Theorem (LCM) does not hold under $\cal{P}$, and because of this result, he considered $\cal{P}$ as a not fruitful topology (for details see \cite{Low_path}). We have a bit of a disagreement on this conclusion, and we will discuss about it in particular, in the next section.

A list of people \footnote{For example, Nada, Agarwal, Shrivastava, Dossena, Williams; for a complete list of names and articles see \cite{Zeemanlike}.} have studied different topologies in the class $\mathfrak{Z}-\mathfrak{G}$, using tools from general topology. There is a little concern about this study: even if it is interesting to know the topological properties of several Zeeman-G\"obel topologies, there is a lack of unity in notation and a common motivation is absent, throughout the existing literature; there are scattered results on whether a separation axiom is satisfied or not, results with respect to connectedness, metrizability, etc., but there is a lack of a main question. The question, in our opinion, should be not to simply find alternative ``better'' \footnote{``Better'' in a topological sense: that is, topologies easier to work with and rich in topological properties.} topologies to the manifold topology $T_M$, but to justify which is the most {\em natural} topology for a spacetime manifold. There is an obvious qualitative difference between the two approaches.

As an example of this general problem, we mention the Fermat Real Line ${}^\bullet\mathbb{R}$, which was defined by Giordano and Kunzinger \footnote{For a short survey, see section 5, from \cite{Nestsandtheirrole}.} as a possible alternative to Synthetic Differential Geometry, aiming to develope new foundations of smooth differential geometry for finite and infinite-dimensional spaces. Two different topologies were introduced on this line, the so-called ``Omega Topology'' and the ``Fermat Topology''; the first topology is generated by a complete metric and is linked to the differentiation of smooth functions on infinitesimals and the latter one 
is generated by a complete pseudo-metric and is linked to the differentiation of non-standard smooth functions. Both topologies play a different role, but none of them is a natural topology for ${}^\bullet\mathbb{R}$; a linearly ordered set should be assigned to its natural topology which is induced by the order. So, it is easy for a confusion about which properties are ``natural'' to appear; for example, continuity properties, under a topology different from the natural topology, might not hold within the natural topology. A simple example which illustrates this issue in spacetime geometry is given by the Zeno sequences, in 
\cite{Zeeman1}.

In the sequence of papers, \cite{Order-Light-Cone}, \cite{On-Two-Zeeman-Topologies}, \cite{LimitCurve}, \cite{Spacelike} and \cite{Minkowskicausal}, the authors aim to establish a common background for the topologisation problem of a spacetime. This background is the Lorentz ``metric'' and the structure of the lightcone, where one can define the chronological order $\ll$, the causal order $\prec$, the relation horismos $\rightarrow$ and also the chorological order $<$; for the last one, see in particular \cite{Spacelike} and for a complete list of relations $R$ depending on the lightcone see \cite{Minkowskicausal}. One can use the following weak version of the {\em interval topology}, in order to get the induced topology from such a relation $R$ on a spacetime $M$.
For a set $X$, consider the sets $I^+(x) = \{y \in X : x R y\}$
and $I^-(x) = \{y \in X : y R x\}$, as well as the collections $\mathcal{S}^+ = \{X\setminus I^-(x) : x \in X\}$
and $\mathcal{S}^- = \{X\setminus I^+(x) : x \in X\}$. A basic-open set $U$ in the weak interval topology $T^{in}$
is defined as $U = A \cap B$, where $A \in \mathcal{S}^+$ and $B \in \mathcal{S}^-$; in other
words, $\mathcal{S}^+ \cup \mathcal{S}^-$ forms a subbase for $T^{in}$. Such topologies where constructed in \cite{Order-Light-Cone}, \cite{On-Two-Zeeman-Topologies} and \cite{Spacelike}, covering the cases of horismos, chronology, causality and chorology (which are lightlike, timelike, causal and spacelike relations, respectively). Such topologies belong to the class $\mathfrak{Z}-\mathfrak{G}$, as we have shown in \cite{Order-Light-Cone}. The seemingly real problem that for each point there exists, for each of these topologies respectively, a local base of unbounded open sets, is solved, by considering the least topology which contains both the manifold topology and a topology $T^{in}$; this topology is called the {\em join topology} or, as it was misnamed by Reed in \cite{Intersection}, the ``intersection topology'' between two given topologies and is defined to be the topology with base $\{U_1 \cap U_2 : U_1 \in T_1 \textrm{ and } U_2 \in T_2\}$, where $T_1,$ $T_2$ are topologies on some set $X$. One can use De Morgan's laws to show that a base for the join topology can be also given by $\{U_1 \cap U_2 : U_1 \in B_1 \textrm{ and } U_2 \in B_2\}$, where $B_1$ is a base for the topology $T_1$ and $B_2$ is a base for the topology $T_2$. In \cite{Spacelike} we have shown that the join topology between $T_M$ and the weak interval topology which is induced by the reflexive chorological order $\leq$ is actually the Path topology of Hawking-King-McCarthy which, in turn, belongs to the class $\mathfrak{Z}-\mathfrak{G}$ and has, locally, an order structure. There is a kind of a dual such topology, studied in \cite{On-Two-Zeeman-Topologies}, which is the join topology between $T_M$ and the reflexive chronological order; this topology, again, has a locally ordered structure.

We now have enough information to dig a bit deeper in the subject, and talk about spacetime singularities.

\section{Singularities, Naked Singularities and a Kind of \\ unexpected Gravitational Time Delay Effects. }

``{{\it{Time stays long enough for anyone who will use it.}}'' - Leonardo da Vinci
\\ \\

In the previous Section we discussed the role of spacetime topology, as a part of the structure of spacetime, and we stressed that, if one sees a spacetime as a mathematical entity, the spacetime topology should be natural. Since the structure of null cone cannot be recovered by the manifold topology \footnote{We refer, again, to \cite{Zeeman1} for a rigorous proof.}, we have excluded the manifold topology as a natural candidate topology for a spacetime. There are more serious issues though, in this discussion, that should not be neglected. For example, the Path topology $\cal{P}$ on a spacetime manifold $M$ is finer than the manifold topology $T_M$, it belongs to the class $\mathfrak{Z}-\mathfrak{G}$, it has locally an order structure that connects it with the lightcone, but the Limit Curve Theorem (LCT) under does not hold under $\cal{P}$ (see \cite{Low_path} and for a further discussion, \cite{LimitCurve}). It is evident that the singularity problem depends on the spacetime topology; one can support this, by looking for example the use of the LCT in basic singularity theorems (see \cite{Limit-Curve-Theorems}, \cite{Noteonlct} as well as \cite{Penrose-1965}). In particular, the LCT, under the manifold topology, states that if 
$\gamma_n$ is a sequence of causal curves, $x_n$ is a point on $\gamma_n$,
for each $n$, and if $x$ is a limit point of $\{x_n\}$, then there is
an endless causal curve $\gamma$, passing through $x$, which is a limit
curve of the sequence $\gamma_n$. The failure of this theorem to hold is very important, because it avoids
basic contradiction arguments that are present in the proofs of  (in our knowledge) all singularity theorems. The fact that the LCT holds under $T_M$ does not make the manifold topology a natural topology though. The failure of LCT to hold under a more proper spacetime topology, like $\cal{P}$ for example, should ring a bell about the appearence of singularities in the basic singularity theorems: do these singularity theorems depend exclusively from the use of the manifold topology? Are they a purely topological effect, that sieges to exist if one considers a more appropriate topology?

The above question has almost certainly a positive answer for classical singularity theorems like  in \cite{Penrose-1965}. This is not so obvious though, at least for the case of naked singularities, if one considers the questions raised by Kip S. Thorne in \cite{Thorne}; the laws of general relativity do not enforce chronology protection: it is easy to find solutions to the Einstein field equation that have closed timelike curves (CTCs - for example, Van Stockum's spacetime, G\"odel's solution of the Einstein equation, etc.).  Physicists have generally dismissed such solutions as unphysical ones and Thorne protests against this attitude \footnote{For more details, and Thorne's arguments, read Section 3, from \cite{Thorne}.}. Here we will copy a very important paragraph in our opinion, from this mentioned paper: {\em  It would be rather surprising to me, if Nature uses one protection mechanism in one situation (e.g. collapsing, spinning bodies), a different one in another situation (e.g. moving cosmic strings) and a third mechanism in a third situation (e.g. the interior of a spinning black hole). More likely, there is one universal mechanism that always does the job, if other mechanisms fail}. We feel that such a ``universal mechanism'' is the topology of the spacetime. For example, exactly as the Path topology $\cal{P}$ prevents a spacetime from satisfying the classical singularity theorems (due to the failure of LCT), in a similar way Low has proved that a spacetime is {\em nakedly singular}, if the space of causal curves is non-Hausdorff (Proposition 3.1, \cite{Low_nakedsing}) as well as the following two Propositions, which bring the discussion about singularities into a purely topological context:

\begin{proposition}\label{Prop1}
For a strongly causal spacetime $M$, the following are equivalent:
\begin{enumerate}

\item $M$ has no geodesically accessible singularities.

\item $M$ is causally pseudoconvex.

\item The space of causal geodesics $C$, of $M$, is Hausdorff.

\end{enumerate}
\end{proposition}

\begin{proposition}\label{Prop2}
A strongly causal spacetime $M$ is globally hyperbolic, iff its space of smooth endless causal curves is Hausdorff.
\end{proposition}

This is really a place that one has to dig a bit deeper; since the Einstein Field Equation permits solutions which bring us in front of CTCs, one has to place the problem of ``rejecting specific solutions as unphysical'' to topology; we are tempted to conjecture that, if there is a final and definite answer about which is the natural topology for a spacetime, then if under such a topology there is no (interior topological) mechanism to avoid CTCs, then one should not have the right reject such solutions with CTCs as unphysical. If, on the other hand, under {\em the} natural topology of a spacetime classical singularities fail to hold, then one has the right to claim that such theorems have no physical meaning.

Here we feel also commenting about the ``in fashion'' technique to increase the spacetime dimensions, in order to ``make the zeros disappear'' (for a discussion, see \cite{Arefourdimensionsenough}). As an example, in \cite{Cotsakis1} and \cite{Cotsakis2} the authors have built a model of a five-dimensional space, whose conformal infinity is our four-dimensional spacetime, its ``ambient boundary''. The aim of this model was to create a topological environment where basic singularity theorems would not hold any longer (see in particular \cite{Cotsakis2} and \cite{Ordr-Ambient-Boundary} as well as \cite{Singularities_on_Amb_B}). The authors finally concluded that the topology on the ambient boundary should be the Fine Zeeman topology $F$; we have corrected this errattum in \cite{Singularities_on_Amb_B}, as the $F$ refers to special relativity while G\"obel's general relativistic analogue $\mathcal{F}$ would be a more appropriate topology to use in a curved spacetime. We have also mentioned that the argument that the ``lack'' of ``Euclidean-open-balls'' does not necessarily imply the lack of singularities is incorrect. First of all, in a curved spacetime an open-ball will be defined via a Riemann metric and not through the natural Euclidean metric. Secondly, since the topologies in the class $\mathfrak{Z}-\mathfrak{G}$ are finer than the manifold topology $T_M$, it is obvious that every open set in $T_M$ will be also open in a topology $T$ in $\mathfrak{Z}-\mathfrak{G}$; such erratta, that are not rare in models in spacetime geometry, show why we need to take methods of general topology more seriously \footnote{For a critical survey on this discussion, we refer to
\cite{withFabio}.}. The authors of \cite{Cotsakis1} and \cite{Cotsakis2} though have had an interesting idea: to sort of ``force'' the ambient boundary, in their model, to be equipped with a topology in Zeeman-G\"obel class, so that the LCT does not hold (that would work with the Path topology $\cal{P}$, for example, as we have already mentioned). And here comes the critical question: why is there a need then to increase the spacetime dimensions, while such a topology would ``hide the infinities'' already in four dimensions?

To bring this discussion a bit further; in  Proposition \ref{Prop1}, there is a connection between pseudoxonvexity and geodesically accessible singularities. 

\begin{definition} 
A spacetime $M$ is {\em causally pseudoconvex} if, for any compact set $K$ in $M$ there exists another compact set $K'$ in $M$, such that any causal geodesic segment with endpoints in $K$ lies in $K'$. 
\end{definition}

A step further from our discussion on singularities will be a discussion on some kind of ``time dilation' phaenomena, in general relativity, which were noticed by Sijie Gao and Robert M. Wald in \cite{Gao-Wald}. We focus our attention in the Theorem below.

\begin{theorem}[Gao-Wald] Let $(M,g_{ab})$ be a null geodesically complete spacetime, satisfying the null energy condition (NEC) and the null generic condition (NGC). Then, given any compact region $K \subset M$, there exists another compact region $K'$ containing $K$, such that if $q,p \notin K'$ and $q \in J^+(p)-I^+(p)$, then any causal curve $\gamma$ connecting $p$ to $q$ cannot intersect the region $K'$.
\end{theorem}

Gao and Wald claim that their Theorem contains some suggestion of a general ``time delay'' phenomena in general relativity, but since $K'$ could be far larger than $K$, it is difficult to make a strong argument for this kind of interpretation of the theorem. In \cite{OnGaoWald}, we have interpreted Gao-Wald Theorem in terms of sliced spaces, and we have shown that $K'$ can be chosen as a ``small enough'' causal diamond containing $K$. There is a more general issue here though: for the proof of Gao-Wald Theorem, the role of the manifold topology $T_M$ is vital. Based on simple topological arguments (see \cite{OnGaoWald}), we see that if one used for example the Path topology $\cal{P}$, or any topology in the class $\mathfrak{Z}-\mathfrak{G}$, the Gao-Wald Theorem will fail to hold, and so some of the corollaries that follow like, for example, the one (Corollary 1 from \cite{Gao-Wald}) which states that there is an absense of particle horizons, in a class of cosmological models, will fail as well.

We believe that the evidence that classical spacetime singularities depending on LCM, naked singularities depending on causal pseudoconvexity and ``time-dilation'' effects of the type of Gao-Wald, are all topological effects is strong, and thus such results are more topological in their nature and ``less physical''. 

\section{A Duality between Timelike-Spacelike Events: between ``Chronos'' and ``Choros''.}

In article \cite{Spacelike} we have studied a duality between two order relations, in Minkowski spacetime $\mathcal{M}$; the chronological order $\ll$ and the ``chorological''   \footnote{Choros stands for space, in Greek, like chronos stands for time.} order $<$, as well as their induced topologies. In order to define these orders, we need to have a closer look to the lightcone of an event $x$ first.

For an event $x \in \mathcal{M}$, we define the following sets:
\begin{enumerate}

\item $C^T(x) = \{y : y=x \textrm{ or } Q(y-x) < 0\}$, the {\em time-cone} of $x$,

\item $C^L (x) = \{y : Q(y-x) = 0\}$, the {\em light-cone} of $x$,

\item $C^S(x) = \{y : y=x \textrm{ or } Q(y-x) > 0\}$, the {\em space-cone} \footnote{Here the word``cone'' is used in a generalised sense, i.e. it is a cone on $I \times \mathbb{S}^{n-2}$ in Minkowski space $\mathcal{M}^n$. } of $x$,

\item $C^{LT}(x) = C^T(x) \cup C^L(x)$ the union of the time- and light-cones of $x$, also known as the {\em causal cone} of $x$, and

\item $C^{LS}(x) = C^S(x) \cup C^L(x)$ the union of the space- and light-cones of $x$.

\end{enumerate}

In \cite{Minkowskicausal} we present all possible relations (to our knowledge), in $\mathcal{M}$ that are related to the Lorentz ``metric'' and their induced topologies. Here we will highlight the following to ones:
$x \ll y$ iff $y \in C_+^T(x)$ ({\em chronology}) and
for non causally-related events $x,y \in M$, $x<y$ iff $y \in C_+^S(x)$, where we have defined $C_+^S(x)$ for some fixed choice of $m \in M$ ({\em chorology}). For a precise and analytical mathematical description of the partition of the spacecone $C^S(x)$ into two spaces, $C_+^S(x)$ and $C_-^S(x)$, we refer to \cite{Spacelike}. Here we will comment on the significance of this duality, without focusing on its technical details. In particular, Zeeman, in \cite{Zeeman1}, stated three alternative topologies to his Fine topology $F$. Several authors, all listed in \cite{Zeemanlike}, have worked on these topologies and in particular in \cite{Order-Light-Cone} and \cite{Spacelike} we have shown that these topologies are join topologies of the Euclidean topology $\mathbb{R}^4$ and a particular weak interval topology; the topology which has a local base of open sets of the form $B_\epsilon(x) \cap C^T(x)$, of bounded timecones (of a radius $\epsilon>0$) by Euclidean balls, is the join of the topology on $\mathbb{R}^4$ and the weak interval topology generated by $<$, while the topology which has a local base of open sets of the form $B_\epsilon(x) \cap C^S(x)$, of bounded spacecones, is the join of the topology on $\mathbb{R}^4$ and a weak interval topology generated by $\ll$. In a few words, we have two topologies in $\mathfrak{Z}-\mathfrak{G}$ (or, to be more precise, in $\mathfrak{Z}$) such that, the one is generated by open sets which are bounded timecones, the other by spacecones and, respectively, the one has locally an order structure by a spacelike (chorological) order while the other (which is generated by bounded spacecones) by a timelike (chronological) order.

We conjecture that this duality exists in curved spacetimes, as well, but one will need to find an alternative root to define a partition of tilted spacecones, to that one that we followed in \cite{Spacelike}, and create a spacelike orientation dual to timelike orientation. We believe that there is strong evidence that this problem is consistent; wherever there is (relativistic) spacetime, there are events, and wherever there are events there are lighcones \footnote{Indeed, there are solutions of the Einstein field equation in general relativity, which imply an extreme tilt of the lightcones which lead for example to CTCs: independently of whether there exists a chronology protection mechanism in a more general frame, something that was conjectured by Hawking, or if such solutions are once accepted (see \cite{Thorne}), we should underline that our discussion lies within the scope of general relativity and not where the theory collapses within a singularity.} and there can be relations depending on the lightcone, such as chronology $\ll$, causality $\prec$ and horismos $\rightarrow$. Since the spacecone is defined in Minkowski space $\mathcal{M}$ as the complement of the causalcone, one has to define general relativistic analogues of the half-planes $P_+(x)$ and $P_-(x)$ that we defined in \cite{Spacelike}. A general relativistic analogue of $<$ will certainly be of a high interest, as one would be able to talk about a duality between timelike and spacelike, in the frame of general relativity, something that might give insights about the passage from locality to nonlocality.

\section{Questions.}

The preceeding four sections raise more questions than to those that are supposed to answer. 

\begin{enumerate}

\item As we mentioned in section 3, the LCT holds under $T_M$ and not under $\cal{P}$. In fact, there is a wider range of topologies within $\mathfrak{Z}-\mathfrak{G}$ where LCT fails to hold while there are other topologies where LCT holds \footnote{See \cite{LimitCurve} for an introductory discussion on this particular problem.}. Roughly speaking, we have topologies which incorporate the causal structure of a spacetime and the classical singularity theorems cannot be formed, while -on the other hand- these singularity theorems are formed when using other topologies, like $T_M$ for example, which do not incorporate the causal structure of the spacetime, but are linked with the metric of the manifold structure. One could probably view this phenomenon from the perspective of Google Earth: depending from the choice available in the package, one could view satellite photos of the Earth in significant detail while, with the use of a different choice, one could make a road system appear, intervening with the satellite picture or, with another choice, one could simply view the civil map of a city with the anaglyph disappearing completely. 

It might be that different topologies reveal a different perspective of spacetime, but is there a topology which is actually the smallest one from all these spacetime topologies that contains all the information that each one of them contain?

\item Given the topologies in the class $\mathfrak{Z}-\mathfrak{G}$, the general relativistic analogue $\mathcal{F}$, of the Fine topology $F$, is incomparable with several of them, including $\cal{P}$; it might be that the condition for a topology to belong to the Zeeman-G\"obel class might exclude topologies which have a significance, and might be appropriate candidates to be called natural topologies. There are such topologies that are mentioned in \cite{Low_nakedsing}, such as the topologies $\mathcal{T}^0$ and $\mathcal{T}^1$, which by themselves belong to a class which contains finer topologies than each of them respectively, which are defined on the space of smooth endless causal curves, in a very natural way, indeed; a further study of these topologies is needed, as they give the topological conditions for a spacetime to be globally hyperbolic (Proposition 4.3, from \cite{Low_nakedsing}) and connect global hyperbolicity to metrizability (Proposition 4.4). 

\item  Given a general relativistic analogue to the partition of the spacecone that we studied in \cite{Spacelike} (which is, still, an open question), it would be interesting to know if the spacelike geodesics form a submanifold, study their topology, as well as their convergence. Given a +-ve spacelike orientation, dual to the timelike orientation, is there a duality in results regarding the space of timelike or causal geodesics with the spacelike ones? A similar question holds for the the space of endless spacelike curves (always under the frame of \cite{Spacelike}), and a possible duality to results concerning the space of causal endless curves. Before attempting any study related to this general question, one should not forget that acausal is a global property while spacelike is a local one.

\item An idea, that was first communicated with the mathematician Santanu Acharjee, is to consider a spacetime as a bitopological space choosing, for example, the manifold topology and another appropriate spacetime topology (for example in the class of $\mathfrak{Z}-\mathfrak{G}$) to serve the definition of bitopological space. It would be interesting to examine if such a topology incorporates the causal, differential, conformal structure of a spacetime and if it is useful to handle with.

\item Kip Thorn's comments, in \cite{Thorne} on rotating contracting bodies and CTCs are linked to the Einstein field equation, and are seeminghly independent from the topology of a spacetime.
In the Low's work, in \cite{Low_nakedsing} it is clear that the naked singularities are a topological effect. How could one connect these two seemingly different results? 

\item Having stated the previous question, on particular solutions to the Einstein field equation leading to CTCs, it is tempting to pose the following related question. In a spacetime maniold, is there a metrizable topology finer than the manifold and coarser than the Fine one? 

\end{enumerate}

There is some criticism about diminiscing returns: why one should continue a study on the topology of a spacetime, if we have not concluded to something general and fruitul yet. We dare to write that such a question is not fruitful, because the topological problems that were mentioned in this Chapter, including the singularity problems that are topological in nature, are too crucial to be ignored.

\section*{Acknowledgments.}
The author wishes to thank Rolf Suabedissen, from Oxford, for being kind to reply to our topological questions, even if they were elementary; we are grateful for his valuable time and for the collegiality. Infinite thanks to Andreas Boukas for sharing thoughts on quantum gravity, some of which we have incorporated in the last section.

\end{document}